\documentstyle[sprocl]{article}
\newcommand{\bm}[1]{\mbox{\boldmath $#1$}}     

\def\PLB{{\em Phys. Lett.}  B}

\def\PRD{{\em Phys. Rev.} D}

\def\GRG{\em Gen. Rel. Grav.}

\def\mco{\multicolumn}

\def\ra{\rightarrow}

\def\ko{K^0}

\def\bea{\begin{eqnarray}}
\def\eea{\end{eqnarray}}

\begin{document}

\title{ON THE (NON) EXISTENCE OF A GRAVITOMAGNETIC DYNAMO }
\author{J.-F. Pascual-S\'anchez}
\address{Dept. Matem\'atica Aplicada Fundamental, 
Secci\'on Facultad de Ciencias,
Universidad de Valladolid, 47011, Valladolid,\\ Spain\\ 
E-mail: jfpascua@maf.uva.es}

\maketitle
\abstracts{Due to the resemblance between Maxwell and the gravitomagnetic equations
obtained in the weak field and slow motion limit of General Relativity,
 one can ask if it is possible to amplify a seed intrinsic rotation or 
 spin motion by a gravitomagetic dynamo, in analogy with the well-known
 dynamo effect.
Using the Galilean limits of the gravitomagnetic equations, the answer 
to this question is negative,due to the fact that a "magnetic" Galilean
limit for the gravitomagnetic equations is physically inconsistent.}

\section{Gravitomagnetic and Maxwell equations}

I am sure that this subject is well-known to the reader, 
so I will try to go quickly in this introductory section.\\

This work is devoted to the possibility of a gravitomagnetic dynamo, 
which would amplify and maintain an initial seed spin motion. 
The starting point is the resemblance between Maxwell and the 
linear and slow motion aproximation of the Einstein's equations 

Hence, I do not start from the full non-linear Einstein's equations, 
to develop, after the projection into the local rest spaces of a 
congruence of observers, the Maxwell analogy in General Relativity, 
based on the correspondence between the Faraday tensor of the electromagnetic
field and the Weyl tensor of the gravitational tidal field. This analogy has been developed in several recent papers, however,
it was put forward and clearly exposed in  the references ~\cite{ma}, ~\cite{bel1}, ~\cite{pen}, ~\cite{el},  ~\cite{ja}.

This approach was done without any aproximation and, in this framework, 
the Bianchi identities are dynamical and the Einstein equations can be 
interpreted  as constitutive relations of a 4-dim non-linear  elastic continuum, (this can be seen in ~\cite{yo}).

As I said, I will follow a more simple path in this work and 
I will start from the linearized Einstein's equations. On this line, 
to begin with,
I will basically use the corresponding chapters of the following
books, ~\cite{oh},~\cite{wa}, with some changes in the notation and new symbols.\\

To center the subject, we begin with several historical remarks, perhaps not always sufficiently reminded.
In the Newtonian theory of gravity, no fundamental gravitational force
is associated with the rotation of a mass. In this theory, if a body rotates,
the gravitational force it exerts on other masses changes only to the 
extent that the matter distribution within the body is affected by the rotation.
The Newtonian gravitational force is only associated to the distribution of
mass at a time, but not with the state of intrinsic rotation of this mass.\\

The development of electrodynamics in the last century and the close analogy
between Coulomb's law of electrostatics and Newton's law of gravity, led to
Holzm\"uller (1870) and Tisserand (1872), to suggest that a gravitational
"magnetic" (or better rotational) force is associated with a rotating mass
due to the "mass current" generated by the rotation, in close analogy
with electrodynamics.\\
Later Heaviside (1893), in an appendix of his book 
{\sl On Electromagnetic Theory} and assuming wave propagation for the gravitational 
field as for the electromagnetic one, wrote Maxwell's equations for
gravitating bodies. Changing in this case, either the sign of the sources
or the sign of the fields, to account for the attractive nature of the
gravitostatic Newtonian field between masses.

Heaviside, in a sucessive paper, applied these Maxwell-like equations
for gravity to the motion of the Sun-Earth system through the cosmic ether,
in order to explain the excess perihelion precession of Mercury
by the new gravitational "magnetic" force. \\

The Mercury perihelion precession was explained several years later by
the Einstein non-linear theory of gravity. Thus, the {\it ad hoc} introduction
of a gravitational "magnetic" field by Heaviside became moot until
Lense and Thirring (1918) and Thirring (1921) showed that, a certain
gravitomagnetic field is indeed associated with the rotation of a mass,
in the framework of the weak field aproximation to General Relativity.\\

Beginning by the Einstein's equations
\begin{equation}\label{1}
R_{ab}-{\textstyle\textstyle{\frac{1}{2}}}g_{ab}\,R=-8\pi\, T_{ab},
\end{equation}
where $a,b = 0,1,2,3$, are space-time indices and we use geometrized units, 
$G = c = 1.$
If the gravity field is weak the metric (gravitational potential) $g_{ab}$, 
can be descomposed as 
\begin{equation}\label{2}
g_{ab}=\eta_{ab}+h_{ab},
\end{equation} 
where $\eta_{ab}={\rm diag}\, (1,-1,-1,-1)$ is the Minkowski metric and $h_{ab}$ is the perturbation due to the weak gravity field. Usually, it is more convenient to use a "new" perturbation $\bar{h}_{ab}$, for calculational purposes and physical reasons. 
\begin{equation}\label{3}
g_{ab}=\eta_{ab}+\bar{h}_{ab}-{\textstyle\frac{1}{2}}\,\eta_{ab}\,\bar{h},
\end{equation}
where
$\bar{h}=\bar{h}^a_a$ is the trace.
Introduce now the object
\begin{equation}\label{6}
G^{abc}={\textstyle\frac{1}{4}}(\bar{h}^{ab,c}-\bar{h}^{ac,b}
 +  \eta^{ab}\,\bar{h}^{cd}_{\;\;\;\; ,d} - \eta^{ac}\,\bar{h}^{bd}_{\;\;\;\; ,d}),
\end{equation}
where $``,´´$ represents the partial derivative. Impose now a Lorentz-like gauge condition in analogy with electrodynamics, which represents the choice of harmonic coordinates:
\begin{equation}\label{7}
\bar{h}^{ab}_{\;\;\;\; ,b}=0.
\end{equation}
From (\ref{6}) and (\ref{7})
\begin{eqnarray}
G^{a[bc]}   &=&G^{abc},\\ \label{8}
G^{[abc]}   &=&0,          \\ \label{9}
G^{d[ab,c]}&=&0,             \label{10}
\end{eqnarray}
where $[\cdot\cdot]$ is the antisymmetry symbol.\\

Putting (\ref{3}) and (\ref{6}) into (\ref{1}) and keeping  only linear terms, one obtains the weak field equations:
\begin{equation}\label{11}
\frac{\partial G^{abc}}{\partial x^c}=-4\pi\, T^{ab} .
\end{equation}
Introduce, now, new symbols
\begin{equation}\label{12}
\begin{array}{llll}
\bm{g}=(g^1,g^2,g^3)\hspace{5mm} & g^i=G^{00i}\hspace{5mm} && i=1,2,3      \\[5mm]
\bm{a}=(a^1,a^2,a^3) & a^i={\textstyle\frac{1}{4}}\,\bar{h}^{0i}  \hspace{5mm}&    &                         \\[5mm]
\bm{b}=(b^1,b^2,b^3) & b^1=G^{023} &                      b^2=G^{031} \hspace{5mm}&  b^3=G^{012}       \\[5mm]
\bm{b}=\nabla\wedge \bm{a} &&& G^{0ij}=a^{i,j}-a^{j,i}                          .
\end{array}
\end{equation}
Put (\ref{12}) into (\ref{10})  and (\ref{11}), one obtains, when the first order effects of the motion of the sources are taken into account, the following Maxwell-like ({\it gravitomagnetic}) equations, invariant under the Poincar\'e (or even Conformal) group.\\
\begin{eqnarray}
\nabla \bm{g}             & =& -4\pi T^{00}= -4\pi\, \rho, 			  \\ \label{13}
\nabla\wedge \bm{b} &=& -4\pi\,\rho\,\bm{u}+							 \displaystyle\frac{\partial {\bm{g}}}{\partial t}, \\ \label{14}
\nabla\wedge \bm{g} &= &-\displaystyle\frac{\partial \bm{b}}{\partial t},\\ \label{15}
\nabla \bm{b}            &=&0. 					     \label{16} 
\end{eqnarray}
Also, from the geodesic equation (equations of motion)
\begin{equation}\label{18}
\frac{d^2x^a}{d\tau^2}+\Gamma^a_{bc}\,\frac{dx^b}{d\tau}\frac{dx^c}{d\tau}=0,
\end{equation}
for a weak stationary field, one obtains the Lorentz-like force law
\begin{equation}\label{19}
\frac{d\bm{u}}{dt} =\bm{g}+ 4\,\bm{u}\wedge \bm{b} ,
\end{equation}
where $\bm{u}$ is the velocity of the test particle.
Note the factor 4 in the gravitomagnetic force term.
Moreover, for a weak stationary field one obtains the gravitomagnetic potential
\begin{equation}\label{20}
\bm{a} =-{\textstyle\frac{1}{2}}\frac{\bm{S} \wedge \bm{r}}{ r^3},
\end{equation}
and the gravitomagnetic field
\begin{equation}\label{21}
\bm{b}=\nabla\wedge \bm{a}=-{\textstyle\frac{1}{2}}\frac{3\bm{n}(\bm{S}\cdot \bm{n})-\bm{S}}{ r^3},
\end{equation}
where $\bm{S}$ is the intrinsic angular momentum of the source and 
$\bm{n}$ is the unit position vector. These equations are analogous 
to the electromagnetic ones, changing the magnetic dipole moment by minus 
twice the angular momentum.
They are used in the GP-B gyroscope and LAGEOS III experiments to obtain 
the precession of a test body due to $\bm{b}$ field.
Finally, note that the electromagnetic analog further iluminates the 
interdependence of the mass-energy currents effects (Lense-Thirring 
precession) and
mass-energy effects (DeSitter-Fokker or geodetic precession), which in
the electric-Galilean limit (will be discussed), are related as
\begin{equation}\label{22}
\bm{b}^\prime=\bm{b}-\bm{v}\wedge \bm{g},
\end{equation}
where $\bm{b}^\prime$ would be the gravitomagnetic field measured, 
for instance, in the proper frame of the gyroscope, $\bm{b}$ would 
be measured in a "fixed" laboratory (at a terrestial Pole, for example) 
and $\bm{v}\wedge \bm{g}$, contributes to the geodetic precession.

\section{Magnetic dynamo theory}
The term "dynamo action or effect" in magnetohydrodynamics (hereafter MHD) 
is generically used to describe the systematic and sustained generation of 
magnetic energy as a result of the stretching action of a velocity field
$\bm{u}$, on a magnetic field $\bm{B}$. In other words, if a conducting 
fluid moves in a magnetic field $\bm{B}$, the flow will be affected by 
the force due to the interaction between $\bm{B}$ and the currents of 
the fluid. Also, $\bm{B}$ will be modified by the currents of the fluid
and this is just the dynamo effect.

We do not write down the whole forbidding set of PDE of MHD, 
whose complexity precludes any hope of a systematic analytic 
or even numerical treatment. Instead, one studies particular 
aspects or makes simplifyng hypotheses.
One of them is the so-called kinematic approach to the dynamo 
effect. In this approach, the velocity field $\bm{u}$ of the fluids 
regarded as known for any time and fixed and the back reaction of 
the magnetic field $\bm{B}$ on $\bm{u}$, via the Lorentz force, is assumed 
negligible. The kinematic dynamo is the most simple case
of self-excited one and considers the evolution (amplification) of the
magnetic field according to the induction equation:
\begin{equation}\label{23}
\frac{\partial \bm{B}}{\partial t}=\nabla\wedge(\bm{u}\wedge \bm{B})+\frac{1}{4\pi}\,\eta_e\, \Delta \bm{B},
\end{equation}
being $\eta_e$, the resistivity or difussivity 
(for insulators is infinite, for plasmas is cero), 
the reciprocal of the electric conductivity $\sigma$.
\begin{equation}\label{}
\eta_e=\frac{1}{\sigma}
\end{equation}
The induction equation is obtained by taking the "macroscopic" 
magnetic Galilean limit (will be discussed) of Maxwell's equations, 
when the displacement current is neglected. 
In this limit the Ampere equation is:
\begin{equation}\label{24}
\nabla\wedge \bm{B}=4\pi\, \bm{J}.
\end{equation}
From this and from the Ohm's law:
\begin{equation}\label{25}
\bm{J}=\sigma(\bm{E}+\bm{u}\wedge \bm{B}),
\end{equation}                         
one can obtain a expression for the electric field $\bm{E}$ and 
substituting it into the Faraday's equation
\begin{equation}\label{26}
\nabla\wedge\bm{E}=-\frac{\partial \bm{B}}{\partial t},
\end{equation}
one obtains the MHD induction equation (\ref {23}).
Of course, we do not will study in this work the usual magnetic dynamo
coupled to the gravitomagnetic field of a Kerr metric.
This very interesting study has been recently initiated and developed 
by Khanna and Camenzind in several papers. They build its formalism 
upon the membrane paradigm for black hole horizons of Damour and 
Thorne \& Macdonald. See for instance,~\cite{ka} ~\cite{nu}. \\

Instead, I will try to propose a similar mechanism to the magnetic dynamo effect in gravitomagnetism, 
to amplify $\bm{b}$ and hence the intrinsic angular momentum $\bm{S}$, due to the fact that we have Maxwell-like equations for gravity at our disposal.
However, the key equation of the kinematic magnetic dynamo, which 
stablish 
the loop to amplify $\bm{B}$ is the Ohm's law. Do we have a similar 
equation in gravity?.

\section{An analog for gravitomagnetism of the Ohm's law}
Our main radical and new idea 
is that, in order to have a gravitomagnetic dynamo, the source fluid 
can not be a perfect fluid. The fluid must be "not dry", wet, and 
hence must have viscosity. But, as viscosity is a tensorial object
and as we need a scalar, we only consider its trace, the viscous pressure,
neglecting the shear viscosity. Viscosity (viscous pressure) $\eta$, will 
be the analog in gravitomagnetism of the resistivity $\eta_e$, for a 
conducting electrical medium.

Our Ohm's-like law for the moving viscous fluid, in a moving frame, 
will be:
\begin{equation}\label{99}
\bm{j}=\rho\, \bm{u}=\delta\left(\bm{g}+\bm{u}\wedge \bm{b}\right),
\end{equation}
where $\bm{j}$ is the mass current that appears in the first term 
of the r.h.s. of (\ref{14}) and being $\delta=1/\eta$, the "dryness"  
of the viscous fluid.

The reason of this analogy is based on the following physical 
picture of the Lense-Thirring effect. Consider, for instance, 
the GP-B gyroscope which will be Fermi transported along his trajectory. 
His precession will be measured by a Frenet-Serret transported frame 
(or fixed with respect to distant stars). This preccesion could be pictured
by considering a spinning sphere immersed in a viscous fluid and a rod
(the axis of the spinning gyroscope). If the rod is placed, at the poles, 
into the viscous fluid, orthogonal to the axis of rotation of the sphere, it
would precess in the same direction as the sphere rotates. However, 
if the rod is placed at the equator, also orthogonal to the axis of rotation
of the sphere, then, it would precess on the opposite direction to the sense 
of rotation of the sphere.

\section{Galilean limits of the gravitomagnetic equations}
It is well-known that Maxwell's equations and the Lorentz force law 
have two different kinds of Galilean limits: electric and magnetic.
This is due, from the mathematical point of view, to the existence 
of two different kinds of Galilean four-vectors.
Starting from a Lorentz four-vector, for instance $(\rho_e,\bm{J})$, 
this can be more timelike, i.e.,  $|\rho_e|  >\!\!>|\bm{J}|$, and in 
this electric Galilean limit, its transformation under the Galilean inertial 
one is
\begin{equation}\label{100}
\rho_e^\prime=\rho_e,\qquad \bm{J}^\prime=\bm{J}-\bm{v}\, \rho_e,
\end{equation}
and for the electric $\bm{E}$ and magnetic field $\bm{B}$ one obtains:
\begin{equation}\label{101}
\bm{E}^\prime=\bm{E},\qquad \bm{B}^\prime=\bm{B}-\bm{v}\wedge \bm{E}.
\end{equation}

Alternatively, the current four-vector can be more spacelike, i.e.,
$|\rho_e|<\!\!< |\bm{J}|$. This case corresponds to the magnetic Galilean 
limit and the corresponding transformation formulas are:
\begin{equation}\label{103}
\rho_e^\prime=\rho_e-\bm{v}\, \bm{J},\qquad \bm{J}^\prime=\bm{J}
\end{equation}
\begin{equation}\label{104}
\bm{E}^\prime=\bm{E}+\bm{v}\wedge \bm{B},\qquad \bm{B}^\prime=\bm{B}.
\end{equation}

Physically, in the electric limit one describes situations where 
isolated electrical charges move at low velocities. On the other hand, 
the magnetic Galilean limit is the usual situation at the macroscopic 
level where magnetic effects are dominant, due to the balance between
negative and positive electric charges.

This magnetic Galilean limit is the proper one that is used in 
magnetic dynamo theory but it is not possible in gravitomagnetism, 
where we do not have negative masses at our disposal. 
Thus, in gravitomagnetism, if we take a Galilean limit, 
this must necessarily be of the electric (almost Newtonian) kind and
describe situations where isolated masses move at low velocities.
In this electric (almost Newtonian) limit, the gravitomagnetic 
equations (\ref{13},\ref{14} and \ref{16}) have the same expressions, 
but there is an important difference, in this limit the Faraday-like 
equation (\ref{15}) has not induction term, this equation now read
\begin{equation} \label{150}
\nabla\wedge \bm{g} =  0,
\end{equation}
And the Lorentz-like force law (\ref{19}), is reduced 
in this electric (almost newtonian) limit, by imposing Galilean 
invariance, to:
\begin{equation}\label{190}
\frac{d\bm{u}}{dt} =\bm{g} ,
\end{equation}
And finally, in this electric (almost Newtonian)-Galilean limit,  
the transformations laws of the fields, under Galilean inertial 
transformations, read as:
\begin{equation}\label{}
\bm{g}^\prime=\bm{g},
\end{equation}
\begin{equation}\label{220}
\bm{b}^\prime=\bm{b}-\bm{v}\wedge \bm{g}.
\end{equation}
Hence the gravitomagnetic field $\bm{b}$ exists, but has no 
observable effects at the level of forces and torques. Moreover, 
{\it in this limit, the proper one for gravity, it is impossible 
to build a gravitomagnetic dynamo even if we have an Ohm's like law
for gravity as (\ref{99})}, because we do not have, at our disposal, 
an induction term in the Faraday's equation.

The only possibility that remains, in my opinion, 
to construct a gravitomagnetic dynamo, would be to consider 
a non-relativistic generalized Newtonian theory of gravity  
of the kind introduced 
by Bel ~\cite{bel3}. This possibility will be explored in a future work.

\section*{Acknowledgments}
I am grateful to A. San Miguel and F. Vicente for discussions and TeX help and, after the completion of this paper, to Ll. Bel for drawing my attention to one of his publications. This work has been partially supported by the project VA61/98
  of Junta de Castilla y Le\'on (Spain).

\end{document}